# Discovery of Magnetic Antiskyrmions Beyond Room Temperature in Tetragonal Heusler Materials

Ajaya K. Nayak,[1] Vivek Kumar,[2] Peter Werner,[1] Eckhard Pippel,[1] Roshnee Sahoo,[2] Françoise Damay,[3] Ulrich K. Rößler,[4] Claudia Felser,[2] Stuart S. P. Parkin[1*]

[1]Max Planck Institute of Microstructure Physics, Weinberg 2, 06120 Halle, Germany
[2]Max Planck Institute for Chemical Physics of Solids, Nöthnitzer Str. 40, 01187 Dresden, Germany
[3]Laboratoire Léon Brillouin, CEA-CNRS, CEA Saclay, 91191 Gif-sur-Yvette, France
[4]IFW Dresden, P.O. Box 270116, 01171 Dresden, Germany

**Skyrmions, topologically stable spin textures, are of great interest for new generations of spintronic devices[1-3]. In general, the stabilization of skyrmions has been achieved in systems with broken inversion symmetry, where the asymmetric Dzyaloshinskii-Moriya interaction (DMI) modifies the uniform magnetic state to a swirling state[4,5]. Depending on the crystal symmetries, two distinct types of skyrmions, Bloch[5,6] and Néel types[7], have been observed experimentally. Here, we present, the experimental manifestation of a special type of spin-swirling, namely *antiskyrmions,* in a family of acentric tetragonal Heusler compounds with $D_{2d}$ crystal symmetry. A spiral magnetic ground-state, which propagates in the tetragonal basal plane, is transformed into a skyrmion lattice-state under magnetic fields applied along the tetragonal axis over a wide temperature interval. Direct imaging by Lorentz Transmission Electron Microscopy (LTEM) shows field stabilized antiskyrmion lattices and isolated antiskyrmions between 100 K and 400 K, and zero-field metastable antiskyrmions at low temperatures.**

Skyrmions are topologically stable vortex-like objects with a swirling spin configuration which have recently been observed in several bulk materials and thin films [4-13]. Skyrmions are potential candidates for new generation spintronic devices[1-3], such as a novel high density, solid-state storage memory device - Racetrack Memory[14-16]. The existence of skyrmions was predicted in extended magnets with acentric crystalline structures more than 25 years ago[17,18]. In these magnets, the magnetization, twisted by relativistic spin-orbit couplings in the magnetic ground-state state, is usually a one-dimensional helix, a *Dzyaloshinskii spiral*. Evidence of the existence of skyrmions emerged from experiments on a small class of chiral magnetic systems with broken inversion symmetry, where they were discovered only in a very narrow temperature-magnetic field region near the magnetic ordering temperature of their magnetic phase diagram [5]. Later the skyrmionic phase was stabilized over a wider temperature range in thin plates of certain chiral magnets [6,8]. Depending upon the spin rotation and by analogy to the two fundamental types of Bloch and Néel domain walls, two distinct types of Bloch and Néel skyrmions have been observed experimentally to date. Here we report, using Lorentz transmission electron microscopy, the observation of a new type of skyrmion, an *antiskyrmion*, in a Mn-Pt-Sn inverse Heusler compound. The anti-skyrmions exist over a wide temperature range even above room temperature both as isolated objects and in ordered lattices.

Figure 1a,b and c illustrate the spin textures of three classes of skyrmions: (a) Bloch skyrmion, (b) anti-skyrmion, and (c) Néel skyrmion. The corresponding helimagnetic structures from which these various skyrmionic states emerge are shown schematically in Fig. 1d,e, and f. Chiral magnets with the cubic B20 crystal structures with *T* symmetry display Bloch type skyrmions [5,6,8-10]. Bloch skyrmions imaged using Lorentz transmission electron microscopy (LTEM) should show a ring type pattern (Fig. 1g). Unlike Bloch skyrmions, a cross-sectional view of the spin arrangement in a Néel skyrmion represents a cycloid, where spins rotate in a plane along the propagation direction (Fig. 1f). In the case of Néel skyrmions viewed in LTEM, the deflected electrons make a closed loop, hence, no intensity modulation is expected (Fig. 1I). Polar magnets with $C_{nv}$ symmetry [7] and multilayer films with interfacial DMI host Néel skyrmions [12,13,19-21].

Antiskyrmions were theoretically predicted in certain tetragonal materials with acentric crystal structures and $D_{2d}$ symmetry [17,18,22,23]. Cross-sectional schematics along four different directions reveal both helicoid and cycloid spin propagations (see Fig. 1b and e) in an antiskyrmion. This unique rotation of the spin is expected to result in a distinct LTEM pattern as shown in Fig. 1h. In the present spin notation, the transmitted electron beam will converge towards the skyrmion center in the vertical direction, whereas, it will diverge in the horizontal direction. Therefore, a single antiskyrmion is expected to form two bright and two dark spots in a LTEM. Although antiskyrmions were predicted some time ago [17,18,22,23] they have not yet been found as isolated objects nor as arrays. Reports in ref. [24] in engineered Co/Pt multilayers of anti-skyrmions are rather simply achiral spin textures without skyrmionic properties. Similarly, complex spin textures in the B20 compound MnGe are simulated to be arrays of skyrmions and anti-skyrmions [25] but the symmetry of this compound only allows for simple skyrmions.

As the antiskyrmions break the cylindrical symmetry and carry a quadrupolar moment of the magnetostatic charges, their properties may differ from those of Bloch and Néel skyrmions. Thus, they are very interesting objects in their own right. In a broader perspective, acentric magnetic carrying skyrmions are of foremost interest for spintronic applications. A key requirement is a material with above room temperature magnetic ordering, the stability of skyrmion states over an extended temperature range, and the ability to integrate films of such materials into magnetic devices. Of especial importance is to control the magnetic anisotropy, which can influence the competition between a spiral magnetic state, skyrmionic textures, and collinear spin-structures. In this regard, magnetic materials with tetragonal crystal structures possess an advantage of unidirectional anisotropy for which they are widely used in spintronics. Indeed, Heusler compounds, $X_2YZ$ (where $X, Y$ are transition metals and $Z$ is a main- group element), are well known for their potential applications in spintronics, especially in tunneling based devices [26]. Mn$_2YZ$ based magnetic Heuslers provide a perfect platform for the design of anisotropic and acentric room-temperature magnets with flexible magnetic configurations [27-29].

The atomic order that forms an 'inverse Heusler' structure in these crystals type is acentric, whereas the higher symmetry in the cubic form does not favor a chiral inhomogeneous DMI. In many of these crystals a cubic to tetragonal transformation "switches on" the chiral inhomogeneous DMI. Recently, a non-collinear spin configuration was observed in the acentric tetragonal Heusler compound Mn$_2$RhSn[30]. Here we have focused on inverse tetragonal Heusler compounds in which we have replaced Rh by the heavier Pt to increase spin-orbit coupling and thereby DMI. However, we find that Mn$_2$PtSn can not be stabilized without the introduction of significant Mn vacancies. A single phase material Mn$_{1.4}$PtSn that crystallizes in an acentric tetragonal structure with space group $I\bar{4}2m$ could be obtained with uniformly distributed Mn vacancies exclusively in the Mn-Pt NaCl type substructure[31]. Additionally, we found that a small amount of Pd that was substituted in place of Pt enhances the tetragonality (supplementary information). Therefore, we focus on Mn$_{1.4}$Pt$_{0.9}$Pd$_{0.1}$Sn in the present study. In this compound the direct magnetic exchange sets the basic magnetic configuration of the Mn moments in the unit cell: there are two distinct Mn sub-lattices, each with their own magnetic moment. The DMI twists this basic magnetic moment configuration along one given propagation direction, thereby giving rise to a long-range helimagnetic structure. However, the crystal structure of this material belongs to a crystal class with **$D_{2d}$** point symmetry. This ensures that the twisting of the ferromagnetic order takes place only in the tetragonal basal plane. For this symmetry, theoretical considerations have shown that a field applied close to the tetragonal c-axis always stabilizes a skyrmion-lattice state, as there is no competing conical spiral state in the magnetic phase-diagram[30]. This is completely different from the acentric cubic helimagnets, where the applied field always favors conical helices propagating along the field direction[5,6]. Here we find from LTEM that indeed there exists a long range helimagnetic structure with a period that is larger than ~100 nm. The skyrmions appear as a lattice with application of magnetic field for temperatures up to 400 K, which is the magnetic ordering temperature for the present material.

The magnetization versus temperature $M(T)$ data measured in a field of 0.1 T for $Mn_{1.4}PtSn$ and $Mn_{1.4}Pt_{0.9}Pd_{0.1}Sn$ are plotted in Fig. 2a. The parent compound $Mn_{1.4}PtSn$ exhibits a Curie temperature, $T_C$, of about 400 K followed by a second transition to a state with a higher magnetization below ~160 K. This suggests that the change in temperature leads to a reorientation of the moments of the Mn sub-lattices. Neutron diffraction studies on powdered samples of $Mn_{1.4}PtSn$ clearly demonstrate a considerable change in the intensity of the (101) and (004) ferromagnetic peaks around the spin-reorientation transition, signifying a change in the magnetic structure at this temperature (inset of Fig. 2a). As expected, $Mn_{1.4}Pt_{0.9}Pd_{0.1}Sn$ displays a similar $T_C$ as that of $Mn_{1.4}PtSn$ due to the equal number of valence electrons [32]. However, the low temperature spin-reorientation transition is decreased to approximately 125 K in the Pd-doped sample.

The bright field transmission electron microscopy (TEM) study shows a martensitic-like twinned microstructure for $Mn_{1.4}Pt_{0.9}Pd_{0.1}Sn$ as shown in Fig. 2b (upper panel). Most interestingly, the adjacent twin platelets orient at an angle of almost 90° to each other: the platelet labelled 1 possesses a [100] orientation, whereas that labelled 2 exhibits a [001] orientation. The corresponding selected area diffraction patterns (SAED) from each individual platelet are shown in the lower panels of Fig 2b. The high-resolution scanning transmission electron microscope (STEM) image taken on the [100] oriented platelet shows alternate bright and dark atoms corresponding to the columns with Mn-Pt(Pd) and Mn-Sn atoms, respectively (Fig. 2c). The enlarged view of the atomic arrangements and the crystal structure of the present system are shown in the left and right insets of Fig. 2c. The overview of the sample taken in focused and underfocused LTEM modes and zero magnetic field are depicted in Fig. 2d and 2e, respectively. Although the focused LTEM image does not show any change in contrast in comparison to the TEM image in Fig. 2c, additional contrast modulation is observed on the platelet with an [001] orientation in case of the underfocused LTEM image. These stripe like features correspond to a helimagnetic structure propagating in the basal plane of the tetragonal crystal structure. The sinusoidal variation of the in-plane magnetization with the helix period of ~135 nm is shown in the inset of Fig. 2e. Application of magnetic field along [001] modifies the helimagnetic stripes into skyrmions (Fig. 2f). It can be mentioned that in the following at all temperatures (except at 100 K) the skyrmion phase was better nucleated when the field was applied at some angle to [001] (see Fig. 4f). Once the skyrmion phase was stabilized with an oblique field the sample was rotated to apply field along [001] direction and a stable antiskyrmion lattice was formed. It is also important to mention that as the temperature approaches $T_C$, the skyrmion phase can be better nucleated even with applying field directly along [001].

For a close look at the nature of the skyrmions in the present system, an underfocused LTEM image of a single skyrmion taken at 0.29 T field applied parallel to [001] direction is displayed in Fig. 3a. Most interestingly, it exhibits two bright spots and two dark spots along [010] and [100] directions, respectively. The modulation of the contrast can be discerned from the appearance of two peaks with a dip in the middle for the line profile taken along [010] (lower inset) and two dips with a peak in the middle for the line profile taken along [100] (upper inset). This picture clearly matches to the expected LTEM pattern as shown in Fig. 1h, signifying the presence of antiskyrmions. A complete reversal of the contrast is observed in the overfocused LTEM image as shown in Fig. 3b. The formation of a hexagonal lattice of antiskyrmions is depicted in Fig. 3c. A slight distortion of the lattice along [010] direction might be related to the presence of a small amount of in-plane field due to a slight misorientation (≈ ±3°) of the sample away from the exact [001] direction. This is because in our thin crystalline foils so-called bend-contours occur when observing the TEM samples at the pole position, which disturbs the visibility of skyrmions. At low in-plane fields the distortion is mostly in the [010] direction as the helix propagates along [100]. Due to the tetragonal $D_{2d}$ symmetry it is expected that the skyrmions lattices undergo characteristic distortions in oblique fields, as their axis remains locked to the tetragonal axis, and the skyrmions are consequently distorted with the perfectly radial core shifting from the centers of the honeycomb lattice cells of the hexagonal densely packed skyrmion lattitce.

A theoretical simulation of the antiskyrmion lattice in a small oblique field is shown in Fig. 3d. It illustrates the skewed appearance of the skyrmions, which leads to a slight rectangular modification of the honeycomb lattice. The effect is easily understood, as the in-plane component of the magnetic field favors and broadens the quarter of the circulating magnetization along the field, at the cost of the opposite part. A complete distortion of the lattice can be seen in Fig 3e, when the magnetic field was applied at an angle of about 20° to [001] by rotating the sample along the [110] direction. In this case the skyrmions appear as elliptical shapes in the Lorentz micrographs with the center of the skyrmions strongly off-centered. The appearance of double-skyrmion-like features results from the projection of the skyrmion tubes that are oriented obliquely to the illuminating electron-beam and to the applied field. As the rotation angle decreases from Fig. 3e to Fig. 3g, the original quadrupole-like internal magnetization distribution of the antiskyrmions is revealed. For rotation angle θ=8° strong bright spots can be seen at the upper half of the antiskyrmions (Fig. 3g), whereas, the strong bright spots appear at the lower half for θ= -5° (Fig. 3h). In both cases the antiskyrmion lattice exhibits a large distortion along [010]. Nearly symmetrical bright and dark spots with an almost hexagonal nature of the lattice can be seen in Fig. 3i, for an applied field of 0.29 T along [001] (within ±3° limit). A small increase in the field to 0.33 T perturbs the regular arrangement of the skyrmions in the lattice (Fig. 3j). This field corresponds to the stability limit of the equilibrium lattice phase. Owing to their topological stability, a great number of antiskyrmions remain as metastable excitations in the homogeneous field-polarized collinear state. For further higher fields antiskyrmions disappear from the relatively thinner region of the sample and only stabilize at the thicker region (Fig. 3k). Finally, the antiskymion lattice evolves into an array of single antiskyrmions, which disappear for fields above 0.49 T at room temperature (Fig. 3l).

Since $Mn_{1.4}Pt_{0.9}Pd_{0.1}Sn$ exhibits a $T_C$ of about 400 K, it is expected that the material should display skyrmions up to 400 K. Underfocused LTEM pattern taken at 350 K in presence of a field of 0.22 T shows a lattice of antiskyrmions (Fig. 4a). At higher temperatures the skyrmion phase can be stabilized at lower fields. Figure 4b shows antiskyrmions at 100 K for a field of 0.33 T applied along [001]. At temperatures below the spin-reorientation transition of the present material, it was difficult to nucleate an antiskyrmion lattice by applying magnetic field even at an angle to [001] after cooling down the sample in zero field. Therefore, the sample was field cooled to 100 K at 0.24 T and subsequently the field was increased or decreased to observe antiskyrmions at different fields (see supplementary information). The antiskyrmion lattice obtained after reducing the field to zero is depicted in Fig. 4c. At zero field each antiskyrmion can be seen with two symmetrical bright spots along [010], whereas, the black spots were smeared out with the background. The zero-field stabilized lattice of antiskyrmions likely is a metastable state, underlining the topological stabilization against the decay into the simple spiral 1D-modulated ground-state.

It is often found that the transition from the helical phase to skyrmions and subsequently to the field polarized state is visualized with the presence of a kink in the $M$ ($H$) measurements [7,33]. $M$ ($H$) isotherms measured at different temperatures for the bulk polycrystalline $Mn_{1.4}Pt_{0.9}Pd_{0.1}Sn$ are shown in Fig. 4d. It can be clearly seen that $M$ ($H$) curves for temperatures down to 150 K exhibit a smeared out (due to polycrystalline nature) kink type of behavior (the region marked by arrows). For temperatures below 100 K an unsaturated magnetic behavior is obtained for fields up to 1 T. The regions marked inside the arrows indicate the phase transition between various magnetic states. $H$-$T$ phase diagrams mainly derived from the LTEM measurements are shown in Fig. 4e. As can be seen the skyrmions can be found for the complete temperature region between 400 K and 100 K (the lowest measurement temperature in the present case). The strength of the magnetic field where the antiskyrmions appear increases with decreasing temperatures. Thus, a large region in the $H$-$T$ phase diagram with stable and possible metastable antiskyrmion-lattice states is realized in the whole temperature region. For temperatures below 200 K the antiskyrmion lattice remains stable even when the field is completely removed. The observation of a stable antiskyrmion phase in zero magnetic field at low temperatures in the field decreasing path can be attributed to the anti-site disorder created by the substitution of Pd in place of Pt in $Mn_{1.4}Pt_{0.9}Pd_{0.1}Sn$. The anti-site disorder act as pinning centers that helps the antiskyrmions to be quenched to a relatively low field when the field is reduced to zero. Since the sample does not exhibit a substantial magnetic hysteresis (see supplementary information) it can be readily considered that the observed effect does not arise from the materials intrinsic hysteretic behavior. The role of disorder is also discussed in the achievement of stable zero field skyrmions in field cooling process in a cubic magnet [34]. However, the stabilization of antiskyrmions at lower fields even without any field cooling/quenching [35] process makes the present system distinct from other skyrmion system. Nevertheless, it is also for the very first time that skyrmions can be stabilized for a temperature up to 400 K in a bulk system.

While we have demonstrated the formation of antiskyrmions in $Mn_{1.4}PtSn$ and $Mn_{1.4}Pt_{0.9}Pd_{0.1}Sn$, we anticipate that the tunable nature of the Heusler family will allow for the formation of a wide range of skyrmionic structures, by varying the number of valence electrons, the spin-orbit coupling, the symmetry of the crystal structure, and thereby the magnetocrystalline anisotropy and the magnetization. In particular, the presence of more than one magnetic sublattice in the family of Heusler materials makes it possible to tune the total magnetic moment to zero [27,28], so as to potentially give rise to antiferromagnetic skyrmions. Such antiferromagnetic skyrmions are predicted to move in a straight line when moved by current in contrast to the curved motion of conventional skyrmions [36]. Interestingly, the quadrupole moment of the magnetostatic charge distribution in the antiskyrmions of acentric magnets allows the control of their motion by applying inhomogeneous magnetic fields.

**References**

1. Jonietz, F. *et al.* Spin Transfer Torques in MnSi at Ultralow Current Densities. *Science* **330**, 1648-1651 (2010).
2. Schulz, T. *et al.* Emergent electrodynamics of skyrmions in a chiral magnet. *Nat. Phys.* **8**, 301-304 (2012).
3. Fert, A., Cros, V. & Sampaio, J. Skyrmions on the track. *Nat. Nano.* **8**, 152-156 (2013).
4. Rossler, U.K., Bogdanov, A.N. & Pfleiderer, C. Spontaneous skyrmion ground states in magnetic metals. *Nature* **442**, 797-801 (2006).
5. Muhlbauer, S. *et al.* Skyrmion Lattice in a Chiral Magnet. *Science* **323**, 915-919 (2009).

6. Yu, X.Z. *et al.* Real-space observation of a two-dimensional skyrmion crystal. *Nature* **465**, 901-904 (2010).
7. Kézsmárki, I. *et al.* Neel-type skyrmion lattice with confined orientation in the polar magnetic semiconductor GaV4S8. *Nat. Mater.* **14**, 1116-1122 (2015).
8. Yu, X.Z. *et al.* Near room-temperature formation of a skyrmion crystal in thin-films of the helimagnet FeGe. *Nat. Mater.* **10**, 106-109 (2011).
9. Seki, S., Yu, X.Z., Ishiwata, S. & Tokura, Y. Observation of Skyrmions in a Multiferroic Material. *Science* **336**, 198-201 (2012).
10. Tokunaga, Y. *et al.* A new class of chiral materials hosting magnetic skyrmions beyond room temperature. *Nat. Commun.* **6**, 7638 (2015).
11. Milde, P. *et al.* Unwinding of a Skyrmion Lattice by Magnetic Monopoles. *Science* **340**, 1076-1080 (2013).
12. Heinze, S. *et al.* Spontaneous atomic-scale magnetic skyrmion lattice in two dimensions. *Nat. Phys.* **7**, 713-718 (2011).
13. Moreau-Luchaire, C. *et al.* Additive interfacial chiral interaction in multilayers for stabilization of small individual skyrmions at room temperature. *Nat. Nano.* **11**, 444-448 (2016).
14. Parkin, S.S.P., Hayashi, M. & Thomas, L. Magnetic domain-wall racetrack memory. *Science* **320**, 190-194 (2008).
15. Yang, S.-H., Ryu, K.-S. & Parkin, S.S.P. Domain-wall velocities of up to 750 ms$^{-1}$ driven by exchange-coupling torque in synthetic antiferromagnets. *Nat. Nano.* **10**, 221-226 (2015).
16. Parkin, S.S.P. & Yang, S.-H. Memory on the Racetrack. *Nat. Nano.* **10**, 195-198 (2015).
17. Bogdanov, A.N. & Yablonsky, D.A. Thermodynamically Stable Vortexes in Magnetically Ordered Crystals - Mixed State of Magnetics. *Sov. Phys. JETP* **68**, 101-103 (1989).
18. Bogdanov, A. & Hubert, A. Thermodynamically stable magnetic vortex states in magnetic crystals. *J. Magn. Magn. Mater.* **138**, 255-269 (1994).
19. Boulle, O. *et al.* Room-temperature chiral magnetic skyrmions in ultrathin magnetic nanostructures. *Nat. Nano.* **11**, 449-454 (2016).
20. Woo, S. *et al.* Observation of room-temperature magnetic skyrmions and their current-driven dynamics in ultrathin metallic ferromagnets. *Nat. Mater.* **15**, 501-506 (2016).
21. Jiang, W. *et al.* Blowing magnetic skyrmion bubbles. *Science* **349**, 283-286 (2015).
22. Bogdanov, A.N., Rößler, U.K., Wolf, M. & Müller, K.H. Magnetic structures and reorientation transitions in noncentrosymmetric uniaxial antiferromagnets. *Phys. Rev. B* **66**, 214410 (2002).
23. Koshibae, W. & Nagaosa, N. Theory of antiskyrmions in magnets. *Nat. Commun.* **7**, 10542 (2016).
24. Zhang, S., Petford-Long, A.K. & Phatak, C. Creation of artificial skyrmions and antiskyrmions by anisotropy engineering. *Sci. Rep.* **6**, 31248 (2016).
25. Tanigaki, T. *et al.* Real-Space Observation of Short-Period Cubic Lattice of Skyrmions in MnGe. *Nano Lett.* **15**, 5438-5442 (2015).
26. Jeong, J. *et al.* Termination layer compensated tunnelling magnetoresistance in ferrimagnetic Heusler compounds with high perpendicular magnetic anisotropy. *Nature Comm.* **7**(2016).
27. Nayak, A.K. *et al.* Design of compensated ferrimagnetic Heusler alloys for giant tunable exchange bias. *Nat. Mater.* **14**, 679-684 (2015).

28. Sahoo, R. *et al.* Compensated Ferrimagnetic Tetragonal Heusler Thin Films for Antiferromagnetic Spintronics. *Adv. Mater.* **28**, 8499–8504 (2016).
29. Nayak, A.K. *et al.* Large Zero-Field Cooled Exchange-Bias in Bulk $Mn_2PtGa$. *Phys. Rev. Lett.* **110**, 127204 (2013).
30. Meshcheriakova, O. *et al.* Large non-collinearity and spin-reorientation in the novel $Mn_2RhSn$ Heusler magnet. *Phys. Rev. Lett.* **113**, 087203 (2014).
31. Jamijansuren, B. *et al.* $Mn_3Pt_2Sn_2$ — A vacancy stabilized Heusler-like compound. (2017).
32. Graf, T., Felser, C. & Parkin, S.S.P. Simple rules for the understanding of Heusler compounds. *Prog. Solid State Chem.* **39**, 1-50 (2011).
33. Li, Y.F. *et al.* Robust Formation of Skyrmions and Topological Hall Effect Anomaly in Epitaxial Thin Films of MnSi. *Phys. Rev. Lett.* **110**, 117202 (2013).
34. Karube, K. *et al.* Robust metastable skyrmions and their triangular-square lattice structural transition in a high-temperature chiral magnet. *Nat. Mater.* **15**, 1237-1242 (2016).
35. Oike, H. *et al.* Interplay between topological and thermodynamic stability in a metastable magnetic skyrmion lattice. *Nat. Phys.* **12**, 62-66 (2016).
36. Barker, J. & Tretiakov, O.A. Static and Dynamical Properties of Antiferromagnetic Skyrmions in the Presence of Applied Current and Temperature. *Phys. Rev. Lett.* **116**, 147203 (2016).

**Acknowledgements:**

We thank Horst Blumtritt and Norbert Schammelt for their help in preparing TEM lamella for the present study. This work was financially supported by the ERC Advanced Grant No. (291472) "Idea Heusler".

**Author Contributions**:

A.K.N., C.F. and S.S.P.P. conceived the original idea for the project. A.K.N. performed Lorentz TEM investigations with help of E.P. and P.W. The bulk materials were synthesized by V.K and A.K.N.. F.D., R.S. carried out the powder neutron diffraction study. A.K.N. and R.S. performed magnetic measurements. A.K.N. and S.S.P.P. wrote the manuscript with substantial contributions from all authors.

**Additional information:**

Supplementary information is available in the online version of the paper. Reprints and permissions information is available online. Correspondence and requests for materials should be addressed to S.S.P.P. (stuart.parkin@mpi-halle.mpg.de).

**Competing financial interests:**

The authors declare no competing financial interests.

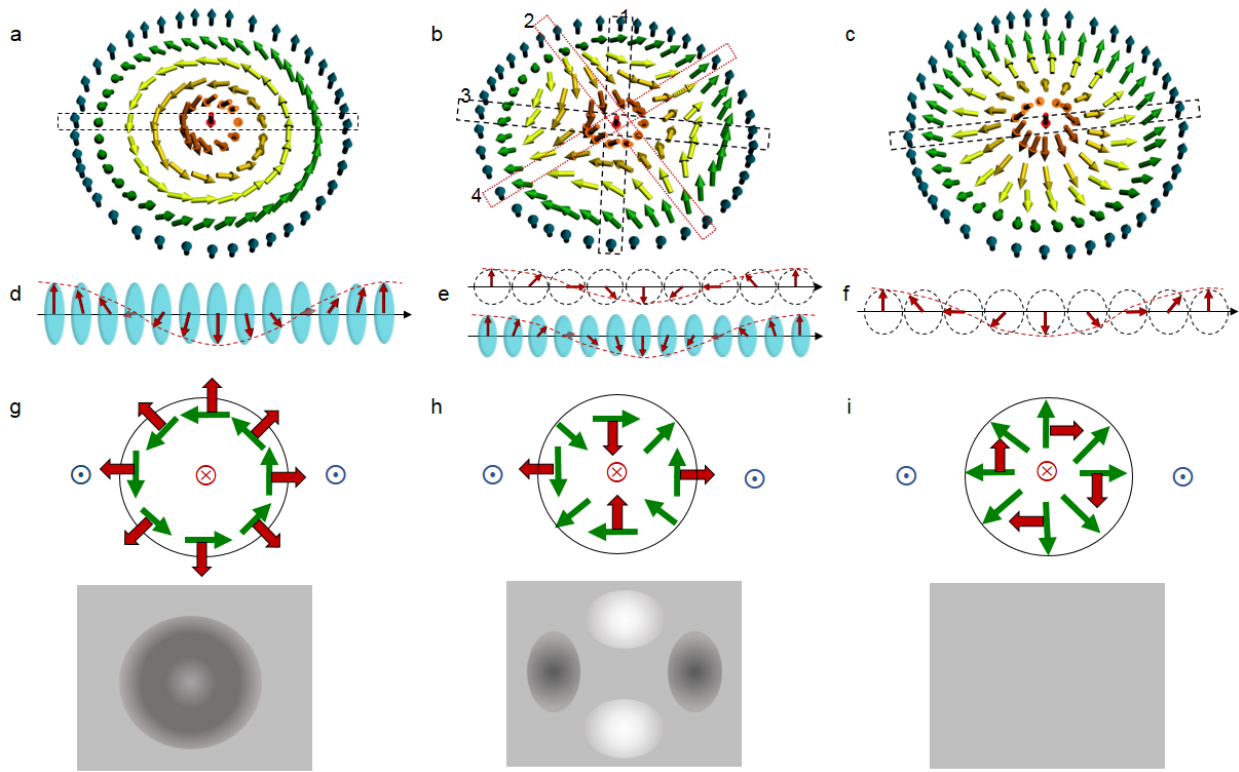

**Figure 1**. **Skyrmions and Antiskyrmions**. Spin textures of **a**, Bloch skyrmion, **b,** antiskyrmion and **c,** Néel skyrmion. In all cases the magnetization is double-twisted along both azimuthal and radial directions. **d,** Cross-section of a Bloch skyrmion along a radial direction (indicated by the dashed rectangle in **a**) corresponds to a transverse helix with an anticlockwise spin rotation (in the present case) whereas **f,** the magnetization is anticlockwise rotated in a spin-cycloid for a Néel

skyrmion. **E,** For an antiskyrmion the magnetization rotates both as a transverse helix (within the dashed rectangles 1and 3 in **b**) and as a cycloid (within the dashed rectangles 2 and 4 in **b**). **g,** Schematic in-plane arrangement of the moments in a Bloch skyrmion (green arrows) and the Lorentz deflections of the transmitted electrons (shown in red arrows) to give a ring type LTEM pattern of the deflected electrons (lower part of the figure). **h,** In-plane spin arrangement of an antiskyrmion. In this case the spins deflect towards the center in the vertical direction, whereas the deflection is outward in the horizontal direction. The corresponding LTEM pattern with two bright spots and two dark spots is exhibited in the lower part of **h**. **i,** In-plane spin arrangement of a Néel skyrmion and the corresponding LTEM pattern.

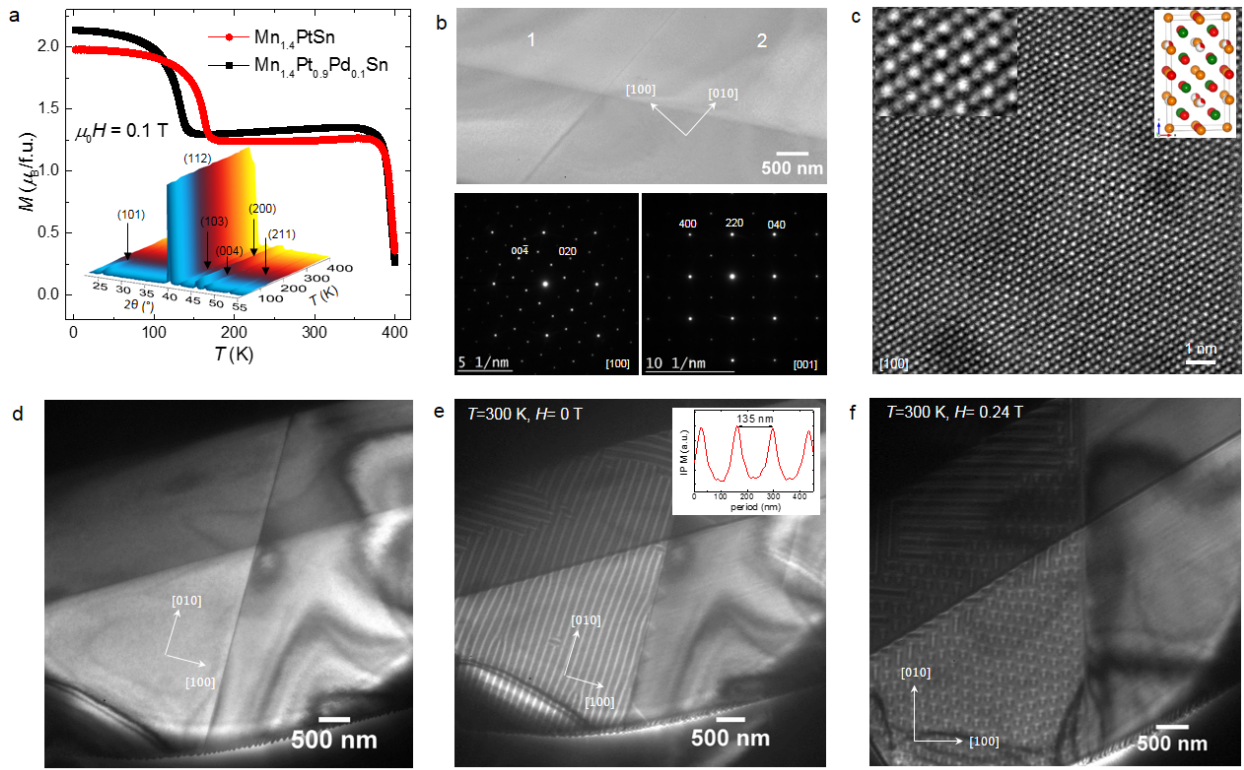

**Figure 2**. **Structural and magnetic properties. a,** Temperature dependence of low-field (0.1 T) magnetization, $M(T)$, for $Mn_{1.4}PtSn$ (filled circles) and $Mn_{1.4}Pt_{0.9}Pd_{0.1}Sn$ (filled squares). Powder neutron diffraction measurements at different temperatures for $Mn_{1.4}PtSn$ are shown in the inset of **a**. **b,** TEM overview of $Mn_{1.4}Pt_{0.9}Pd_{0.1}Sn$ showing a twinned microstructure. Platelet-1 exhibits a crystal orientation [100], whereas, platelet-2 shows [001] orientation. Selected area diffraction patterns (SAED) taken on platelets 1 and 2 are shown in the left and right lower panels, respectively. **c,** High resolution scanning transmission electron microscope (STEM) image taken on the [100] oriented platelet. The left inset shows the enlarged view of the atomic arrangement, whereas the right inset displays the crystal structure of the present Mn-Pt(Pd)-Sn system. **d,** In-focus LTEM image showing twinned microstructure as seen in **c.** Room temperature under-focused LTEM image taken at **e,** zero field and **f,** in a field of 0.24 T, demonstrating the existence

of helical magnetic structure and skyrmions, respectively. The insets of **e** depicts the sinusoidal change of the in-plane magnetization with period of the helix.

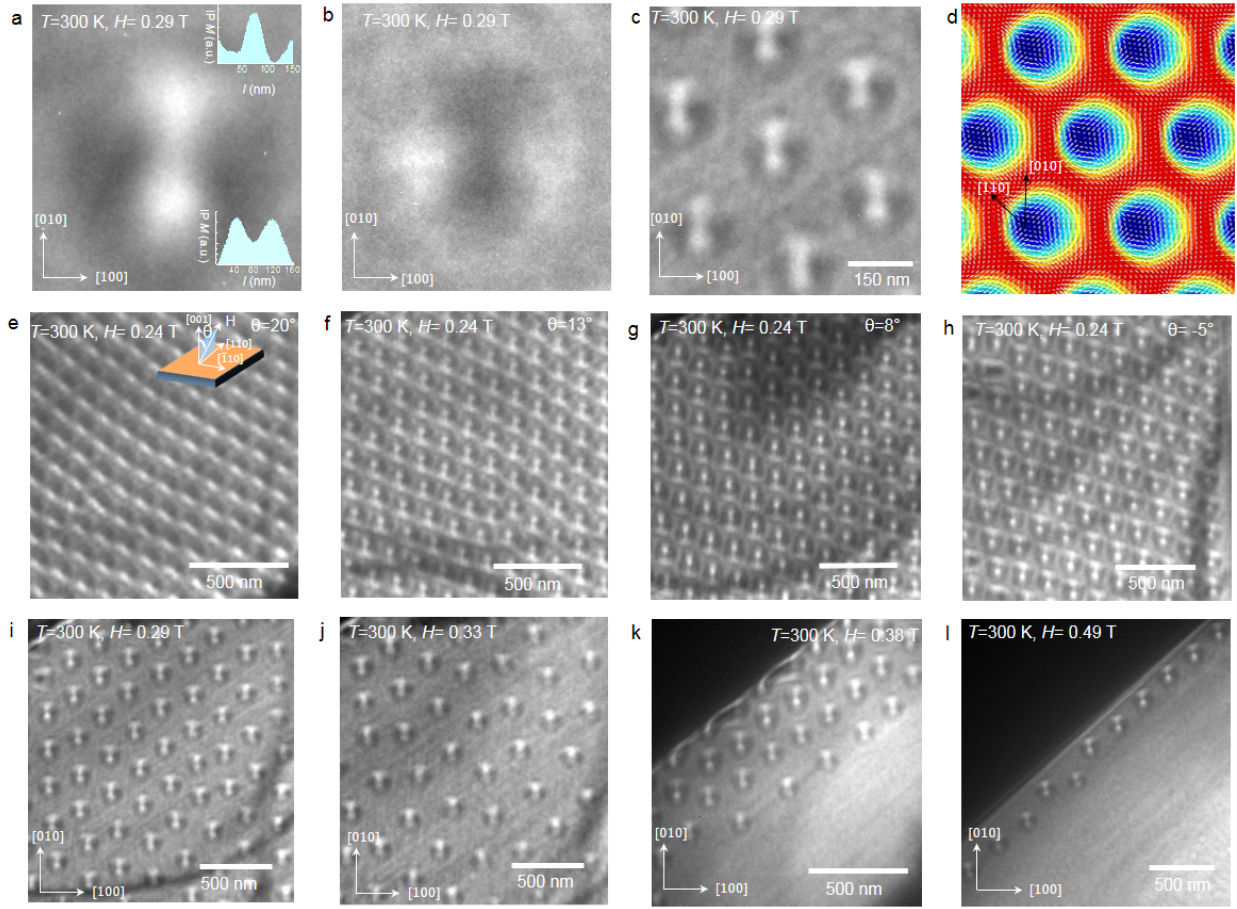

**Figure 3. Room temperature skyrmions in Mn$_{1.4}$Pt$_{0.9}$Pd$_{0.1}$Sn**. **a,** Under-focused LTEM image of a single antiskyrmion with field applied along [001] direction. The lower and upper insets show the intensity profiles of the in-plane magnetization along [010] and [100] directions. **b,** Over-focused LTEM image of the single antiskyrmion shown in **a**. **c,** Under-focused LTEM showing a hexagonal lattice of antiskyrmions. **d,** Theoretical calculation of an antiskyrmion lattice in an oblique field. **e-h,** Under-focused LTEM images of antiskyrmions taken in 0.24 T with different rotation angle (θ) as shown by schematic diagram in the inset of **e**. **i-l,** Under-focused LTEM images of antiskyrmions taken at different fields applied along [001].

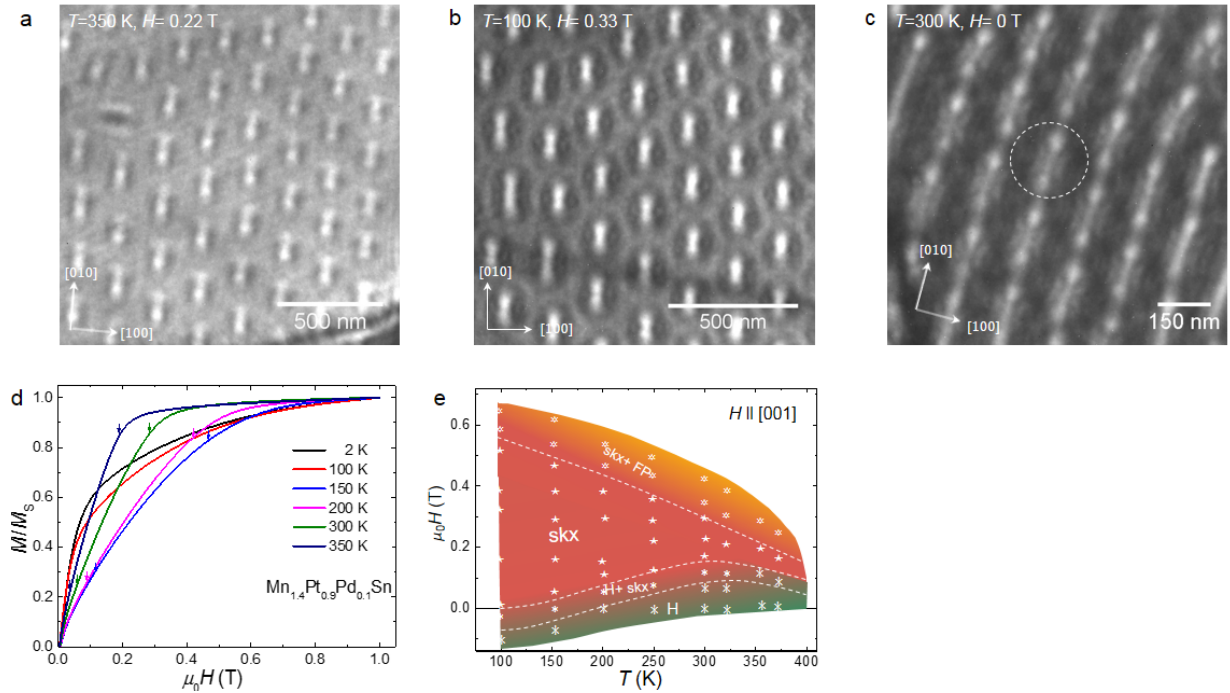

**Figure 4. Temperature dependence of antiskyrmions and phase diagram.** Under-focused LTEM image taken at **a,** 350 K in a field of 0.22 T, **b,** 100 K and field of 0.33 T and **c,** 100 K and zero field. A single antiskyrmion in zero field is shown inside a dashed circle. **d,** Magnetic isotherms, $M(H)$, at different temperatures for $Mn_{1.4}Pt_{0.9}Pd_{0.1}Sn$. The data are normalized by dividing the magnetization value at 1 T. **e,** $H$-$T$ phase diagrams for $Mn_{1.4}PtSn$ derived from the LTEM measurements with field along [001]. The phase diagram represents different phases at different field and temperature values, helical phase (H, ∗), skyrmions (skx, ∗), mixed phase of helical and skyrmions (H+skx, ∗), field polarized state (FP) and mixed phase of skyrmions and field polarized (skx+FP, ∗).